\documentclass[conference]{IEEEtran}

\usepackage{textcomp}
\usepackage{stfloats}
\usepackage{csquotes}
\usepackage{setspace}
\usepackage{graphicx,amsmath,amsfonts,amsthm,color,comment}
\usepackage{enumitem}
\usepackage{enumerate}
\usepackage{algorithmic}
\usepackage{pgfplots}
\usepackage{pgf}
\usepackage{tikz}
\usetikzlibrary{positioning,shapes}
\usepackage{amssymb}
\usepackage{csquotes}
\usepackage{bbm}
\usepackage{multirow}
\usepackage{bm}
\usepackage[nolist]{acronym}
\usepackage{amsthm}
\usepackage{arydshln}
\usepackage{subcaption}
\theoremstyle{definition}

\DeclareMathOperator*{\argmax}{arg\,max}

\usepackage[ruled,linesnumbered,vlined]{algorithm2e}
\SetNlSty{bfseries}{\color{black}}{}
\SetArgSty{textnormal}
\pgfplotsset{compat=1.15}
\SetKw{Continue}{continue}
\SetKw{Break}{break}

\def\ben{\begin{eqnarray*}}
\def\een{\end{eqnarray*}}

\def  \H5G{H_\text{CA-Polar}}

\newcommand*{\affmark}[1][*]{\textsuperscript{#1}}

\begin{document}


\title{Near-Optimal Generalized Decoding \\ of Polar-like Codes}

\author{
	\IEEEauthorblockN{Peihong Yuan\affmark[1], Ken R. Duffy\affmark[2], Muriel M\'edard\affmark[1]}\\
	\IEEEauthorblockA{\affmark[1]\textit{Research Laboratory of Electronics, Massachusetts Institute of Technology} \\   
	\affmark[2]\textit{College of Engineering and College of Science, Northeastern University}\\
	\{phyuan, medard\}@mit.edu, k.duffy@northeastern.edu}
}

\begin{acronym}
    \acro{BCH}{Bose-Chaudhuri-Hocquenghem}
    \acro{GRAND}{guessing random additive noise decoding}
    \acro{SGRAND}{soft GRAND}
    \acro{DSGRAND}{discretized soft GRAND}
    \acro{SRGRAND}{symbol reliability GRAND}
    \acro{ORBGRAND}{ordered reliability bits GRAND}
    \acro{5G}{the $5$-th generation wireless system}
    \acro{APP}{a-posteriori probability}
    \acro{ARQ}{automated repeat request}
    \acro{ASK}{amplitude-shift keying}
    \acro{AWGN}{additive white Gaussian noise}
    \acro{B-DMC}{binary-input discrete memoryless channel}
    \acro{BEC}{binary erasure channel}
    \acro{BER}{bit error rate}
    \acro{biAWGN}{binary-input additive white Gaussian noise}
    \acro{BLER}{block error rate}
    \acro{UER}{undetected error rate}
    \acro{LER}{list error rate}
    \acro{bpcu}{bits per channel use}
    \acro{BPSK}{binary phase-shift keying}
    \acro{BSC}{binary symmetric channel}
    \acro{BSS}{binary symmetric source}
    \acro{CDF}{cumulative distribution function}
    \acro{CSI}{channel state information}
    \acro{CRC}{cyclic redundancy check}
    \acro{DE}{density evolution}
    \acro{DMC}{discrete memoryless channel}
    \acro{DMS}{discrete memoryless source}
    \acro{BMS}{binary input memoryless symmetric}
	\acro{eMBB}{enhanced mobile broadband}
	\acro{FER}{frame error rate}
	\acro{uFER}{undetected frame error rate}
	\acro{FHT}{fast Hadamard transform}
	\acro{GF}{Galois field}
	\acro{HARQ}{hybrid automated repeat request}
	\acro{i.i.d.}{independent and identically distributed}
	\acro{LDPC}{low-density parity-check}
	\acro{LHS}{left hand side}
	\acro{LLR}{log-likelihood ratio}
	\acro{MAP}{maximum-a-posteriori}
	\acro{MC}{Monte Carlo}
	\acro{ML}{maximum-likelihood}
	\acro{PDF}{probability density function}
	\acro{PMF}{probability mass function}
	\acro{QAM}{quadrature amplitude modulation}
	\acro{QPSK}{quadrature phase-shift keying}
	\acro{RCU}{random-coding union}
	\acro{RHS}{right hand side}
	\acro{RM}{Reed-Muller}
	\acro{RV}{random variable}
	\acro{RS}{Reed–Solomon}
	\acro{SCL}{successive cancellation list}
	\acro{SE}{spectral efficiency}
	\acro{SNR}{signal-to-noise ratio}
	\acro{UB}{union bound}
	\acro{BP}{belief propagation}
	\acro{NR}{new radio}
	\acro{CA-SCL}{CRC-assisted successive cancellation list}
	\acro{DP}{dynamic programming}
	\acro{URLLC}{ultra-reliable low-latency communication}
    \acro{MAC}{multiple access channel}
    \acro{SIC}{successive interference cancellation}
    \acro{RLNC}{random linear network coding}
    \acro{SINR}{signal-to-interference-plus-noise ratio}
    \acro{MDR}{misdetection rate}
    \acro{SC}{successive cancellation}
    \acro{SCL}{successive cancellation list}
    \acro{PM}{path metric}
    \acro{SISO}{soft-input soft-output}
    \acro{SO}{soft-output}
    \acro{SOGRAND}{soft-output GRAND}
    \acro{RM}{Reed-Muller}
    \acro{PM}{path metric}
    \acro{BCJR}{Bahl, Cocke, Jelinek and Raviv}
    \acro{MIMO}{multi-input multi-output}
    \acro{BICM}{bit-interleaved coded modulation}
    \acro{BICM-ID}{bit-interleaved coded modulation with iterative decoding}
    \acro{TUB}{truncated union bound}
\end{acronym}

\maketitle

\begin{abstract}
We present a framework that can exploit the tradeoff between the undetected error rate (UER) and block error rate (BLER) of polar-like codes. It is compatible with all successive cancellation (SC)-based decoding methods and relies on a novel approximation that we call \emph{codebook probability}. This approximation is based on an auxiliary distribution that mimics the dynamics of decoding algorithms following an SC decoding schedule. Simulation results demonstrates that, in the case of SC list (SCL) decoding, the proposed framework outperforms the state-of-art approximations from Forney's generalized decoding rule for polar-like codes with dynamic frozen bits. In addition, dynamic Reed-Muller (RM) codes using the proposed generalized decoding significantly outperform CRC-concatenated polar codes decoded using SCL in both BLER and UER. Finally, we briefly discuss three potential applications of the approximated codebook probability: coded pilot-free channel estimation; bitwise soft-output decoding; and improved turbo product decoding.
\end{abstract}

\begin{IEEEkeywords}
Polar coding, generalized decoding, error detection.
\end{IEEEkeywords}
\markboth
	{}
	{}

\section{Introduction}
Reliability in physical layer communication hinges on the frequency of forward error correction decoding errors. Undetected errors occur when the decoder provides a codeword that is distinct from the transmitted one and the system remains unaware of this erroneous decision. Undetected errors can be more harmful than detected errors, which are usually labeled as erasures. Consequently, code and decoder design objectives encompass not only reducing the \ac{BLER} but also maintaining a low \ac{UER}.

Decoding algorithms can be divided into two classes: complete and incomplete. Complete decoders always return valid codewords and any \ac{ML} decoding algorithm essentially belongs to this group. In contrast, an incomplete decoder may provide estimates not fulfilling the conditions of being a member of the underlying code~\cite[Ch.1]{blahut2003algebraic}. If that occurs, the receiver is able to detect the error and so can, for example, ask for a retransmission. Both the \ac{BCJR}~\cite{bahl1974optimal} and \ac{BP} decoding algorithms are well-established incomplete decoders as a result of their focus on making bit-wise decisions for the coded bits.

The standard practical method to convert a complete decoder into an incomplete one is to employ a \ac{CRC} outer code, which provides error-detection capability after using a complete decoder for the inner code. The addition of the \ac{CRC} results in a reduced code-rate, and so the \ac{CRC} should be carefully designed to optimize the trade-off between the \ac{BLER} and \ac{UER}. The notion of an optimal incomplete decoding algorithm, introduced in~\cite{forney1968exponential}, can be viewed as implementing \ac{ML} decoding followed by a post-decoding threshold test that determines whether to accept or reject the \ac{ML} decision. This approach is optimal in the sense that there is no other decoding rule that simultaneously gives a lower \ac{BLER} and a lower \ac{UER}. The metric for evaluating this test can be efficiently carried out for terminated convolutional codes~\cite{raghavan1998reliability,hof2009optimal} and well approximated for tail-biting convolutional codes~\cite{hof2010performance} via a modification to the \ac{BCJR} algorithm.

CRC-concatenated polar codes, as described in~\cite{TV-list,niu2012crc}, refer to the serial concatenation of polar codes with outer CRC codes. \Ac{SCL} decoding~\cite{TV-list} is typically used to decode CRC-concatenated polar codes. First, a \ac{SCL} decoder creates a list of candidate decodings based on the inner polar code. If none of the candidates in the list pass the CRC test, a decoding failure is declared, i.e, an error is detected. Otherwise, the most likely candidate in the list is selected as the final decision, leading to an undetected error if it is not the same as the transmitted message. 

Recently, a soft-output measure has been developed for all soft-input \ac{GRAND} algorithms~\cite{galligan2023upgrade}, which takes the form of an accurate estimate of the \ac{APP} that a single decoding is correct or, in the case of list decoding, the probability that each element of the list is the transmitted codeword or the codeword is not contained in the list. Core to the accuracy of the measure is the approximation that all unidentified codewords are uniformly distributed amongst unexplored sequences.

In this study, we explore generalized decoding approaches for polar~\cite{Arikan09} and polar-like codes, leveraging a \ac{SC}-based decoding algorithm and a novel post-decoding threshold test. By extending the main idea in~\cite{galligan2023upgrade} from a guessing-based search to \ac{SC}-based tree search, we introduce an approximation to a quantity we call the codebook probability, which is the the sum of the probability of all valid codewords given soft-input. Our results demonstrate the ratio between the probability of the decoder output and the codebook probability closely aligns with the probability that \emph{the output decision is correct} for polar-like codes with dynamic frozen bits. The merit of the decoder output is then assessed by comparing the previously mentioned ratio with a threshold. Simulation results show that decoding polar-like codes with dynamic frozen bits and the proposed generalized decoding results in significantly better performance than CRC-concatenated polar codes using \ac{SCL} in terms of both \ac{BLER} and \ac{UER}. We explore potential applications of the codebook probability in coded pilot-free channel estimation, bitwise soft-output decoding, and turbo product code decoding.

This paper is organized as follows. Section~\ref{sec:prelim} gives background on the problem. An approximation of the codebook probability of polar-like codes is proposed in section~\ref{sec:proposed}. Section~\ref{sec:numerical} presents numerical results demonstrating the accuracy of \ac{BLER} prediction based our approximation and the performance of joint error correction and detection. In section~\ref{sec:siso}, we illustrate some other potential applications. Section~\ref{sec:conclusions} concludes the paper.

\section{Preliminaries}\label{sec:prelim}
\subsection{Notations}
In this paper, length-$N$ vectors are denoted as $x^N = \left(x_1, x_2, \dots, x_N\right)$, where we write $x_i$ for its $i$-th entry. For completeness, note that $x^0$ is void. A \ac{RV} is denoted by an uppercase letter, such as $X$, and its counterpart, e.g., $x$, is used for a realization. Then, a random vector is expressed as $X^N = \left(X_1, X_2, \dots, X_N\right)$. The \ac{PDF} of a continuous \ac{RV} and the \ac{PMF} of a discrete \ac{RV} evaluated at $x$ are denoted as $p_X(x)$, where the extensions to the vectors is trivial. A \ac{B-DMC} is characterized by conditional probabilities $p_{Y|C}$, where the input takes on values in binary alphabet $\{0,1\}$ and the output set $\mathcal{Y}$ is specified by the considered channel model. For natural numbers, we write $[a] = \{i : i \in \mathbb{N}, 1 \leq i \leq a\}$.


\subsection{Polar-like Codes and Their Decoding}\label{sec:polar}
A binary polar-like code of block length $N$ and dimension $K$ is defined by a set $\mathcal{A}\subseteq \left[N\right]$ of indices with $|\mathcal{A}|=K$ and a set of linear functions $f_i$, $i\in\mathcal{F}$, where $N$ is a positive-integer power of $2$ and $\mathcal{F}\triangleq\left[N\right]\setminus \mathcal{A}$. The $K$-bit message is mapped onto the subvector $u_\mathcal{A}$ of the input $u_1^N$ to the polar transform, where the frozen bits are evaluated as
\begin{align}
u_i=f_i\left(u^{i-1}\right), \forall i\in\mathcal{F}.\label{eq:dyn_frozen_def}
\end{align}
Observe that each frozen bit $u_i$ is either statically set to zero (since $f_i$ are linear) or they change according to the input $u^{i-1}$, which are called dynamic frozen bits. This representation unifies various modifications of polar codes, e.g., \ac{CRC}-concatenated polar codes~\cite{TV-list}, polar subcodes~\cite{trifonov2015polar} polarization-adjusted convolutional codes~\cite{arikan2019sequential} and dynamic \ac{RM} codes~\cite{cocskun2020successive}. The codeword is then obtained by applying polar transform as $c^N=u^N\mathbb{F}^{\otimes \log_2 N}$, where $\mathbb{F}$ is the binary Hadamard matrix~\cite{Arikan09}. The codeword $c^N$ is transmitted via $N$ independent uses of a \ac{B-DMC}.

At the receiver side, 
\ac{SC} decoding observes the channel output $y^N$ and performs a sequential greedy search to obtain decisions as
\begin{align}
\hat{u}_i=\left\{\begin{aligned}
& f_i\left(\hat{u}^{i-1}\right),~&i\in\mathcal{F}\\
& \argmax_{u\in\left\{0,1\right\}}Q_{U_i|Y^NU^{i-1}}\left(u|y^N\hat{u}^{i-1}\right),~&i\in\mathcal{A} \label{eq:SC_decoding}
\end{aligned}\right.
\end{align}
where $Q_{U^N|Y^N}$ denotes an auxiliary conditional \ac{PMF} induced by assuming that $U^N$ is uniformly distributed in $\{0,1\}^N$. This implies that $Q_{U^N|Y^N}$ assumes the frozen bits $U_\mathcal{F}$ to be also uniformly distributed and independent of the information bits $U_{\mathcal{A}}$. Observe that \ac{SC} decoding computes $Q_{U_i|Y^NU^{i-1}}\left(u_i|y^N\hat{u}^{i-1}\right)$ by treating $U_i$ and all upcoming frozen bits, namely $u_{i+1},\dots,u_N$, as uniformly distributed given the channel observation $y^N$ and previous decisions $\hat{u}^{i-1}$. In other words, the frozen constraints are used to determine which decision to make but don't impact the reliability of the decision. Then, a block error is declared only if $u_{\mathcal{A}}\neq \hat{u}_{\mathcal{A}}$ since Eq.~\eqref{eq:dyn_frozen_def} is already used in decoding via Eq.~\eqref{eq:SC_decoding}.

\ac{SC} decoding, with a slight modification, has the ability to provide soft information (in the form of a probability) associated to any partial input $\hat{u}^i\in\{0,1\}^i$, which is also called a decoding path. In particular, for any given decoding path $\hat{u}^{i}$, one uses the Bayes' rule to write
\begin{align*}
&Q_{U^i|Y^N}\left(\hat{u}^i\left|y^N\right.\right)\nonumber\\
&=Q_{U^{i-1}|Y^N}\left(\hat{u}^{i-1}\left|y^N\right.\right)Q_{U_i|Y^NU^{i-1}}\left(\hat{u}_i\left|y^N\hat{u}^{i-1}\right.\right)
\end{align*}
where the right-most term $Q_{U_i|Y^NU^{i-1}}\left(\hat{u}_i\left|y^N\hat{u}^{i-1}\right.\right)$ can be computed efficiently by the standard \ac{SC} decoding and $Q_{U^0|Y^N}\left(\varnothing|y^N\right)\triangleq 1$ by definition.\footnote{The term $-\log Q_{U^i|Y^N}\left(\hat{u}^i\left|y^N\right.\right)$ is call \ac{PM} for SC-based decoding in \ac{LLR} domain~\cite{balatsoukas2015llr}.} Note that the frozen constraints are used to determine which decision to make at frozen positions behaving as anchors and are irrelevant to the reliability of the decoding path.



\subsection{Forney's Generalized Decoding}\label{sec:Gdecoding}
Forney introduced a generalized decoding rule~\cite{forney1968exponential}, which relies on a threshold test. The decoder output $\hat{c}^N$ is accepted if 
\begin{align}\label{eq:forney}
\frac{p_{Y^N|C^N}\left(y^N\left|\hat{c}^N\right.\right)}{\sum_{c^N\in \mathcal{C}} p_{Y^N|C^N}\left(y^N\left|c^N\right.\right) } \geq \frac{2^{NT}}{1+2^{NT}}
\end{align}
where the threshold parameter $T\geq 0$ controls the tradeoff between \ac{BLER} and \ac{UER}. Otherwise, the decision is rejected and decoder outputs an error flag, resulting in a detected error. Forney's generalized decoding rule is optimal in the sense of minimizing the \ac{UER} for a given \ac{BLER}
(and vice versa). Since the denominator of Eq.~\eqref{eq:forney} is difficult to compute in general, we may use a suboptimal decoding rule~\cite{forney1968exponential,hof2010performance,sauter2023error} based on list decoding with 
\begin{align*}\label{eq:forney_approx}
\!\!\!\frac{p_{Y^N|C^N}\left(y^N\left|\hat{c}^N\right.\right)}{\sum_{c^N\in \mathcal{C}} p_{Y^N|C^N}\left(y^N\left|c^N\right.\right) }\approx \frac{p_{Y^N|C^N}\left(y^N\left|\hat{c}^N\right.\right)}{\sum_{c^N\in \mathcal{L}_C} p_{Y^N|C^N}\left(y^N\left|c^N\right.\right) }.
\end{align*}
where $\mathcal{L}_C$ contains the candidate decisions of codeword $c^N$ obtained from the list decoding. Obviously, the approximation is precise when the list decoder exhaustively enumerates the entire codebook.
 
\section{Generalized Decoding of Polar-like Codes}\label{sec:proposed}

In the following, decision regions for SC-based decoding rely on a threshold test of function $\Gamma\left(y^N, \hat{u}^N\right)$,
\begin{align}
\Gamma\left(y^N, \hat{u}^N\right)
\triangleq\frac{Q_{U^N|Y^N}\left(\hat{u}^N\left|y^N\right.\right)}{\sum_{u^N\in \mathcal{U}} Q_{U^N|Y^N}\left(u^N\left|y^N\right.\right) }
\end{align}
where the set $\mathcal{U}$ contains all decoding path $u^N$ that satisfies the frozen constraints, i.e., 
\begin{align*}
\mathcal{U}\triangleq \left\{u^N\in\left\{0,1\right\}^N: u_i=f_i\left(u^{i-1}\right), \forall i\in\mathcal{F} \right\}.
\end{align*}
As explained in section~\ref{sec:polar}, the probability of the decoding path $u^N$ is irrelevant to the frozen constraints, i.e., $Q_{U^N|Y^N}\left(u^N|y^N\right)$ is not necessarily equal to $0$ even if $u^N$ does not fulfill the frozen constraints.

The core problem of generalized decoding is computing the \emph{codebook probability}, i.e., the sum of probabilities for all valid decoding paths.,
\begin{align}\label{eq:codebook_probability}
Q_\mathcal{U}\left(y^N\right)\triangleq \sum_{u^N\in \mathcal{U}}Q_{U^N|Y^N}\left(u^N|y^N\right).
\end{align}
In this section, we introduce a simple method to approximate the reliability of unvisited codewords based on SC-based decoding of polar and polar-like codes.

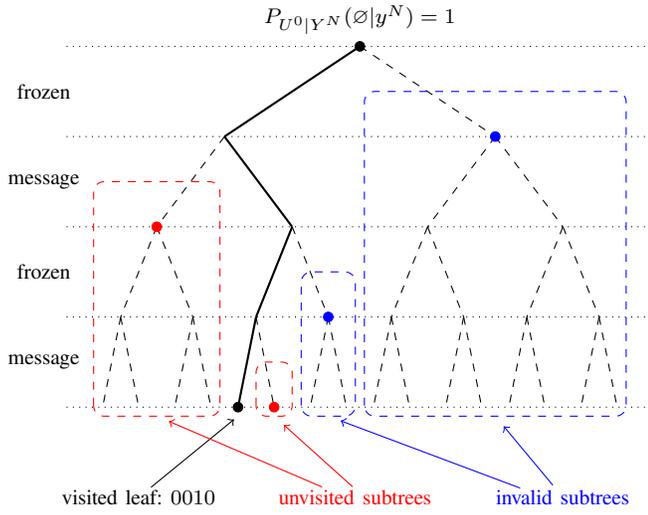
\begin{figure}[t]
	\centering
	\begin{tikzpicture}[scale=1.2]
    \footnotesize
\draw [dashed] (0.5,4)--(0.1,3); \draw [dashed]  (0.5,4)--(0.9,3);
\draw [thick] (2,4)--(1.6,3); \draw [dashed] (2,4)--(2.4,3);
\draw [dashed] (3.5,4)--(3.1,3); \draw [dashed] (3.5,4)--(3.9,3);
\draw [dashed] (5,4)--(4.6,3); \draw [dashed] (5,4)--(5.4,3);

\draw [dashed] (1.25,5)--(0.5,4); \draw [thick]  (1.25,5)--(2,4);
\draw [dashed] (4.25,5)--(3.5,4); \draw [dashed] (4.25,5)--(5,4); 

\draw [thick] (2.75,6)--(1.25,5); \draw[dashed] (2.75,6)--(4.25,5);

\draw[dotted] (-0.5,2) to (6,2);
\node at (-0.75,2.5) {message}; 
\draw[dotted] (-0.5,3) to (6,3);
\node at (-0.75,3.5) {frozen}; 
\draw[dotted] (-0.5,4) to (6,4);
\node at (-0.75,4.5) {message}; 
\draw[dotted] (-0.5,5) to (6,5);
\node at (-0.75,5.5) {frozen}; 
\draw[dotted] (-0.5,6) to (6,6);

\draw [dashed] (0.1,3)--(-0.1,2); \draw [dashed]  (0.1,3)--(0.3,2);
\draw [dashed] (0.9,3)--(0.7,2); \draw [dashed]  (0.9,3)--(1.1,2);
\draw [thick] (1.6,3)--(1.4,2); \draw [dashed]  (1.6,3)--(1.8,2);
\draw [dashed] (2.4,3)--(2.2,2); \draw [dashed]  (2.4,3)--(2.6,2);
\draw [dashed] (3.1,3)--(2.9,2); \draw [dashed]  (3.1,3)--(3.3,2);
\draw [dashed] (3.9,3)--(3.7,2); \draw [dashed]  (3.9,3)--(4.1,2);
\draw [dashed] (4.6,3)--(4.4,2); \draw [dashed]  (4.6,3)--(4.8,2);
\draw [dashed] (5.4,3)--(5.2,2); \draw [dashed]  (5.4,3)--(5.6,2);
\node at (2.75,6.3) {$P_{U^0|Y^N}(\varnothing|y^N)=1$};
\draw[fill] (2.75,6) circle (1.5pt);
\node at (0.3,1) {visited leaf: $0010$};
\draw[fill] (1.4,2) circle (1.5pt);
\draw [shorten >=2pt,shorten <=6pt,->] ((0.3,1)--(1.4,1.9);

\draw[blue, dashed, rounded corners] (2.8,1.9) rectangle (5.7,5.5);
\draw[blue, fill] (4.25,5) circle (1.5pt);
\draw[blue, dashed, rounded corners] (2.1,1.9) rectangle (2.7,3.5);
\draw[blue, fill] (2.4,3) circle (1.5pt);

\draw[red, dashed, rounded corners] (1.6,1.9) rectangle (2.0,2.5);
\draw[red, fill] (1.8,2) circle (1.5pt);
\draw[red, dashed, rounded corners] (-0.2,1.9) rectangle (1.2,4.5);
\draw[red, fill] (0.5,4) circle (1.5pt);

\node at (5,1) {\textcolor{blue}{invalid subtrees}};
\draw [blue,shorten >=5pt,shorten <=6pt,->] ((5,1)--(4.25,1.9);
\draw [blue,shorten >=5pt,shorten <=10pt,->] ((4.8,1)--(2.4,1.9);

\node at (2.7,1) {\textcolor{red}{unvisited subtrees}};
\draw [red,shorten >=5pt,shorten <=6pt,->] ((2.8,1)--(1.8,1.9);
\draw [red,shorten >=5pt,shorten <=10pt,->] ((2.6,1)--(0.5,1.9);

\end{tikzpicture}
	\caption{Example of the SC decoding tree of a polar code with frozen bits $u_1=u_3=0$. The whole decoding tree consists of three parts: a) visited leaf: the SC output $\hat{u}^4=(0,1,0,0)$. b) invalid subtrees: the subtree rooted at $\hat{u}_1=1$ and the subtree rooted at $\hat{u}^3=(0,1,1)$. c) unvisited subtrees: the subtree rooted at $\hat{u}^2=(0,0)$ and the leaf $\hat{u}^4=(0,1,0,1)$.}
	\label{fig:sc_tree}
\end{figure}

An SC-based decoding algorithm divides the decoding tree into three parts,
\begin{itemize}
    \item [a)] \emph{Visited leaves} denote the valid visited decoding paths of depth $N$. Note that we may have more than one visited leaf, e.g., a \ac{SCL} decoder~\cite{TV-list} returns $L$ visited leaves.
    \item [b)] \emph{Invalid subtrees} stands for the subtrees rooted at the nodes which are not visited during the decoding due to the conflict of frozen constraints.
    \item [c)] \emph{Unvisited subtrees} are the subtrees rooted at the nodes which satisfy the frozen constraints, but are not visited (usually due to the complexity issues).
\end{itemize}
Now we define sets $\mathcal{V}$, $\mathcal{W}$ and $\mathcal{I}$ containing the visited leaves, the roots of unvisited subtrees, and the roots of invalid subtrees, respectively. For the mini example shown in Fig.~\ref{fig:sc_tree}, we have 
\begin{align*}
\mathcal{V} &= \left\{(0,1,0,0)\right\}\\
\mathcal{W} &= \left\{(0,0),(0,1,0,1)\right\}\\
\mathcal{I} &= \left\{(1),(0,1,1)\right\}
\end{align*}
The codebook probability $Q_\mathcal{U}\left(y^N\right)$ is then written as 
\begin{align}\label{eq:3set}
\underbrace{\sum_{u^N\in \mathcal{V}}Q_{U^N|Y^N}\left(u^N|y^N\right)}_{\text{(a) all visited leaves}}+\overbrace{\sum_{a^i\in \mathcal{W}} \!\!\underbrace{\sum_{\substack{u^N\in \mathcal{U}\\u^i=a^i}}Q_{U^N|Y^N}\left(u^N|y^N\right)}_{\text{(c) all valid leaves underneath node $a^i$}}}^{\text{(b) all unvisited valid leaves}}.
\end{align}
The term (c) in Eq.~\eqref{eq:3set} describes the sum the of probabilities for all valid decoding paths underneath node $a^i$. We assuming that the leaves are uniformly distributed underneath the unvisited node $a^i$. By extending the approach~\cite[Cor.3]{galligan2023upgrade} from guess-based search to tree search, we use the approximation 
\begin{align}\label{eq:core_approx}
\text{term (c) in Eq.~\eqref{eq:3set}}\approx 2^{-\left|\mathcal{F}^{(i:N)}\right| }P_{U^i|Y^N}\left(a^i|y^N\right)
\end{align}
where $\mathcal{F}^{(i:N)}$ denotes the set of indices for the frozen bits in the future, i.e., 
\begin{align*}
\mathcal{F}^{(i:N)} = \left\{j: j\in\mathcal{F}, i\leq j\leq N\right\}.
\end{align*}
The codebook probability Eq.~\eqref{eq:codebook_probability} is then approximated by
\begin{align}\label{eq:APU}
Q_\mathcal{U}^*\left(y^N\right) &\triangleq \underbrace{\sum_{u^N\in \mathcal{V}} Q_{U^N|Y^N}\left(u^N|y^N\right)}_{\substack{\text{sum of prob. for all visited leaves}}}\nonumber\\
&+\underbrace{\sum_{a^i\in \mathcal{W}} \!\! 2^{-\left|\mathcal{F}^{(i:N)}\right| }Q_{U^i|Y^N}\left(a^i|y^N\right)}_{\substack{\text{approx. sum of prob. for all unvisited valid leaves}}}.
\end{align}

Our algorithm to compute $Q_\mathcal{U}^*\left(y^N\right)$ is compatible with all SC-based decoders. During SC-based decoding, when the decoder decides not to visit a subtree rooted at the node $a^i$, for $i\in\mathcal{A}$, we accumulate the probability $2^{-\left|\mathcal{F}^{(i:N)}\right|}Q_{U^i|Y^N}\left(a^i|y^N\right)$ as the approximated sum of probabilities for all valid leaves underneath node $a^i$.\footnote{The probability $Q_{U^i|Y^N}\left(a^i|y^N\right)$ is computed by the SC-based decoder and does not require extra complexity.}

\section{Numerical Results}\label{sec:numerical}
In this section, we present numerical results demonstrating the accuracy of the approximated codebook probability and the performance of joint error correction and detection. We consider polar-like codes with two types of information sets,
\ac{RM} codes~\cite{muller1954application,reed1954class} and 5G polar codes~\cite{5gpolar}, 
and two types of frozen constraints, static frozen bits with $u_i=0,~i\in\mathcal{F}$, and (convolutional) dynamic frozen bits~\cite{arikan2019sequential},
    \begin{align*}
    u_i=u_{i-2}\oplus u_{i-3}\oplus u_{i-5}\oplus u_{i-6},~i\in\mathcal{F},i>6.
    \end{align*}

\subsection{Accuracy of codebook probability}\label{sec:accuracy_block}
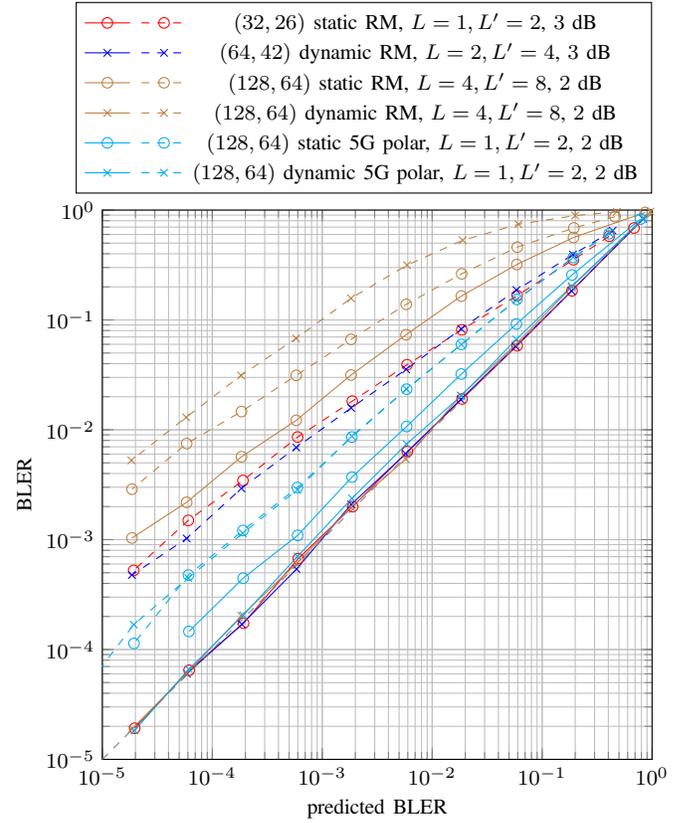
\begin{figure}[t]
	\centering
	\begin{tikzpicture}[scale=1]
\footnotesize
\begin{loglogaxis}[
legend style={at={(1,1.03)},anchor= south east},
legend columns=2,
ymin=1e-5,
ymax=1,
width=3.5in,
height=3.5in,
grid=both,
xmin = 1e-5,
xmax = 1,
xlabel = {predicted BLER},
ylabel = {BLER},
]

\addplot[red,mark=o]
table[]{x y
0.68465 0.6862
0.18701 0.18418
0.059331 0.05821
0.01871 0.019093
0.0059424 0.0063431
0.0018932 0.002
0.00060471 0.00067245
0.00019266 0.00017354
6.1536e-05 6.4731e-05
1.9694e-05 1.9219e-05
};\addlegendentry{ }

\addplot[red,mark=o,dashed,mark options=solid]
table[]{x y
0.40717 0.57757
0.19278 0.35342
0.05926 0.16681
0.018616 0.081329
0.0059093 0.039094
0.0018762 0.018236
0.00059766 0.0085912
0.00019001 0.0034576
6.0495e-05 0.0014962
1.9302e-05 0.00052545
};\addlegendentry{$(32,26)$ static RM, $L=1,L^\prime=2$, $3$~dB}

\addplot[blue,mark=x]
table[]{x y
0.82841 0.83096
0.18517 0.18345
0.057298 0.057451
0.018139 0.018902
0.0057459 0.0060413
0.0018368 0.0020973
0.00058421 0.00053989
0.00018582 0.00016962
5.8782e-05 6.0295e-05
};\addlegendentry{ }

\addplot[blue,mark=x,dashed,mark options=solid]
table[]{x y
0.43791 0.65001
0.19093 0.3937
0.058424 0.18695
0.018358 0.082943
0.0057986 0.035497
0.0018322 0.015809
0.00058147 0.0069023
0.00018406 0.0029298
5.8519e-05 0.001025
1.8584e-05 0.0004769
};\addlegendentry{$(64,42)$ dynamic RM, $L=2,L^\prime=4$, $3$~dB}

\addplot[brown,mark=o]
table[]{x y
0.86614 0.95106
0.1929 0.56292
0.058839 0.3197
0.018424 0.16486
0.0057879 0.073356
0.0018291 0.031521
0.00058014 0.012192
0.00018452 0.0056852
5.8469e-05 0.0021875
1.8601e-05 0.0010356
};\addlegendentry{ }

\addplot[brown,mark=o,dashed,mark options=solid]
table[]{x y
0.46093 0.8709
0.19454 0.68739
0.059393 0.45953
0.018563 0.26302
0.0058086 0.13872
0.0018291 0.066569
0.00058016 0.03129
0.00018431 0.014665
5.8491e-05 0.0074959
1.8606e-05 0.0028797
};\addlegendentry{$(128,64)$ static RM, $L=4,L^\prime=8$, $2$~dB}

\addplot[brown,mark=x]
table[]{x y
0.96441 0.96751
0.18575 0.19933
0.057426 0.059787
0.018176 0.020496
0.0057436 0.005419
0.001827 0.0019762
0.00057762 0.0005991
0.00018355 0.0002025
5.8241e-05 6.0244e-05
1.8451e-05 1.8775e-05
};\addlegendentry{ }

\addplot[brown,mark=x,dashed,mark options=solid]
table[]{x y
0.47733 0.9645
0.19976 0.89098
0.060849 0.74245
0.018929 0.53091
0.0058994 0.31508
0.0018271 0.15733
0.00057904 0.067932
0.00018272 0.031228
5.788e-05 0.012988
1.837e-05 0.0052696
};\addlegendentry{$(128,64)$ dynamic RM, $L=4,L^\prime=8$, $2$~dB}

\addplot[cyan,mark=o]
table[]{x y
0.78461 0.82842
0.18693 0.25671
0.058366 0.091953
0.018415 0.032156
0.005852 0.010721
0.0018652 0.003713
0.00059798 0.0010943
0.00019101 0.00044612
6.1345e-05 0.00014616
};\addlegendentry{ }

\addplot[cyan,mark=o,dashed,mark options=solid]
table[]{x y
0.4073 0.60965
0.19224 0.37051
0.058939 0.15377
0.018446 0.05985
0.0058588 0.02343
0.0018616 0.0085916
0.00059404 0.0029832
0.00018988 0.0012113
6.0738e-05 0.00047601
1.9423e-05 0.00011367
};\addlegendentry{$(128,64)$ static 5G polar, $L=1,L^\prime=2$, $2$~dB}

\addplot[cyan,mark=x]
table[]{x y
0.82749 0.83664
0.18583 0.20332
0.058206 0.066571
0.018397 0.020398
0.0058511 0.0074201
0.0018632 0.0023738
0.00059616 0.00070268
0.00019062 0.00020486
6.1184e-05 6.6146e-05
1.9661e-05 1.8353e-05
};\addlegendentry{ }

\addplot[cyan,mark=x,dashed,mark options=solid]
table[]{x y
0.40756 0.61687
0.19231 0.37312
0.058666 0.15181
0.018449 0.059871
0.0058375 0.023478
0.0018611 0.0087486
0.00059417 0.0028404
0.00018948 0.001144
6.0467e-05 0.00044771
1.9326e-05 0.00016855
3.6577e-06 1.8914e-05
};\addlegendentry{$(128,64)$ dynamic 5G polar, $L=1,L^\prime=2$, $2$~dB}

\addplot[gray,dashed]
table[]{x y
1e-5 1e-5
1 1
};

\end{loglogaxis}

\end{tikzpicture}
	\caption{Predicted \ac{BLER} vs. simulated \ac{BLER} of polar-like codes with proposed scheme and Forney's approximation. The proposed method (solid) works with SCL decoding of list size $L$, while the Forney's approximation (dashed) works with list size $L^\prime$.}
	\label{fig:misdetection_rates_log}
\end{figure}

We use the following metric to evaluate the probability of \emph{the output decision being the transmitted codeword}, 
\begin{align}\label{eq:Gamma}
\Gamma^*\left(y^N, \hat{u}^N\right)
\triangleq \frac{Q_{U^N|Y^N}\left(\hat{u}^N|y^N\right)}{Q_\mathcal{U}^*\left(y^N\right)}.
\end{align}
Fig.~\ref{fig:misdetection_rates_log} plots the \ac{BLER} given the predicted \ac{BLER}. In the \ac{MC} simulation, the codewords are transmitted over \ac{biAWGN} channels. The \ac{SCL} decoder outputs a decision $\hat{u}^N$ and block soft output $\Gamma^*\left(y^N, \hat{u}^N\right)$. We gather blocks with $1-\Gamma^*\left(y^N, \hat{u}^N\right)$ within specific ranges,  
\begin{align*}
\left[1,10^{-0.5}\right),\left[10^{-0.5},10^{-1}\right),\dots,\left[10^{-4.5},10^{-5}\right)
\end{align*}
and compare their \ac{BLER} to the average predicted \ac{BLER} $\text{E}\left[1-\Gamma^*\left(y^N, \hat{u}^N\right)\right]$ (solid lines). For reference, we also show the Forney-style approximation (dashed lines) with list size $\left|\mathcal{L}_U\right|=L^\prime$
\begin{align}\label{eq:Gamma_forney}
\Gamma^\prime\left(y^N, \hat{u}^N\right)=\frac{Q_{U^N|Y^N}\left(\hat{u}^N|y^N\right)}{\sum_{u^N\in \mathcal{L}_U} Q_{U^N|Y^N}\left(u^N|y^N\right) }
\end{align}
where $\mathcal{L}_U$ denotes the list of candidate decisions of $u^N$.\footnote{In this work,  $\mathcal{L}_U$ is associated with the list of candidate decisions for $u^N$, while $\mathcal{L}_C$ is associated with the list of candidate decisions for $c^N$, i.e., $\mathcal{L}_C\triangleq \left\{c^N\in\left\{0,1\right\}^N: c^N=u^N\mathbb{F}^{\otimes \log_2 N}, \forall u^N\in\mathcal{L}_U \right\}$.} Eq.~\eqref{eq:Gamma_forney} approximates the codebook probability $Q_\mathcal{U}\left(y^N\right)$ as the list probability $\sum_{u^N\in \mathcal{L}_U}Q_{U^N|Y^N}\left(u^N|y^N\right)$.

The results in Fig.~\ref{fig:misdetection_rates_log} show that the approximation in~Eq.~\eqref{eq:APU} accurately predicts the \ac{BLER} of the polar-like codes with dynamic frozen constraints. The approximated codebook probability Eq.~\eqref{eq:APU} relies on the assumption of uniform leaf distribution under the unvisited node. However, the static frozen bits (after the first message bit) may disrupt this assumption. Thus, the approximation in~Eq.~\eqref{eq:core_approx} has a mismatch for polar-like codes with static frozen bits. Observe that our approximation yields accurate predictions for static $(32,26)$ RM codes because the count of frozen bits following the first message bit is minimal and insufficient to disrupt the assumption of a uniform leaf distribution.\footnote{There are only $3$ frozen bits after the first message bit for the $(32,26)$ static RM code.} In fact, our approach provides an accurate approximation for codebook probability if polar-like codes with \emph{random} dynamic frozen bits~\cite[Definition 1]{cocskun2020successive} of any length and rate.

\subsection{Joint error correction and detection}
We apply a threshold test,
\begin{align}\label{eq:decision_rule}
\Gamma^*\left(y^N, \hat{u}^N\right) > 1-\epsilon.
\end{align}
The final decision $\hat{u}^N$ is accepted when Eq.~\eqref{eq:decision_rule} is satisfied; otherwise, the decoder returns an error flag. As $\Gamma^*\left(y^N, \hat{u}^N\right)$ evaluates the probability of the output decision being correct, the decision rule mentioned above imposes an upper limit of $\epsilon$ on the \ac{MDR}, where \ac{MDR} is defined as the probability of an accepted decision being erroneous, i.e., the ratio between \ac{UER} and \ac{BLER}.

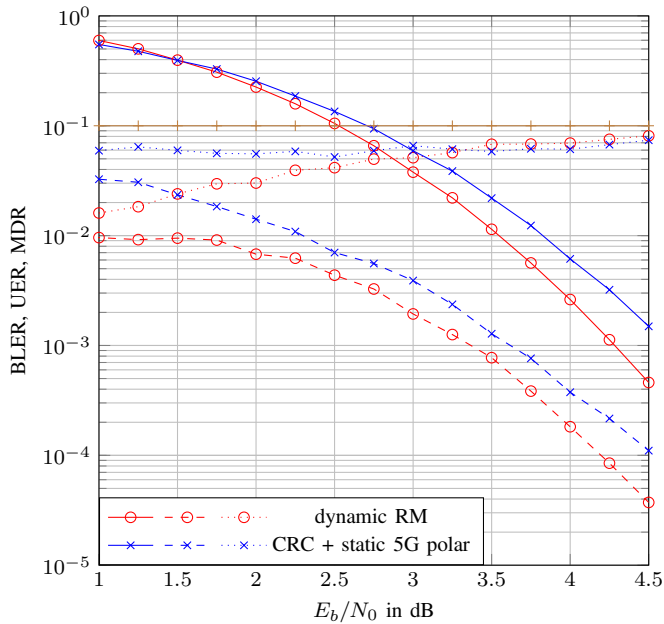
\begin{figure}[t]
	\centering
	\begin{tikzpicture}[scale=1]
\footnotesize
\begin{semilogyaxis}[
legend style={at={(0,0)},anchor= south west},
legend columns=3,
ymin=1e-5,
ymax=1,
width=3.5in,
height=3.5in,
grid=both,
xmin = 1,
xmax = 4.5,
xlabel = $E_b/N_0$ in dB,
ylabel = {BLER, UER, MDR},
]

\addplot[red,mark=o]
table[]{x y
0 0.89347
0.25 0.84035
0.5 0.77716
0.75 0.68958
1 0.59722
1.25 0.5014
1.5 0.39563
1.75 0.30754
2 0.22512
2.25 0.1586
2.5 0.10505
2.75 0.065764
3 0.037813
3.25 0.022068
3.5 0.011418
3.75 0.0056408
4 0.0026207
4.25 0.0011277
4.5 0.00046044
};\addlegendentry{ }

\addplot[red,mark=o,dashed,mark options=solid]
table[]{x y
0 0.0035265
0.25 0.0050878
0.5 0.0064712
0.75 0.0089214
1 0.0095923
1.25 0.0091941
1.5 0.0094841
1.75 0.0091083
2 0.0067776
2.25 0.0062331
2.5 0.0043633
2.75 0.0032678
3 0.0019367
3.25 0.0012571
3.5 0.00077463
3.75 0.00038438
4 0.00018212
4.25 8.4854e-05
4.5 3.7283e-05
};\addlegendentry{ }

\addplot[red,mark=o,dotted,mark options=solid]
table[]{x y
0 0.003947
0.25 0.0060544
0.5 0.0083267
0.75 0.012937
1 0.016062
1.25 0.018337
1.5 0.023972
1.75 0.029616
2 0.030107
2.25 0.0393
2.5 0.041537
2.75 0.049689
3 0.051216
3.25 0.056964
3.5 0.067843
3.75 0.068143
4 0.069493
4.25 0.075245
4.5 0.080972
};\addlegendentry{dynamic RM}

\addplot[blue,mark=x]
table[]{x y
1 0.5493
1.25 0.4763
1.5 0.3934
1.75 0.32815
2 0.25418
2.25 0.18636
2.5 0.13501
2.75 0.09399
3 0.059187
3.25 0.038679
3.5 0.021938
3.75 0.012357
4 0.0061404
4.25 0.0032128
4.5 0.0014933
4.75 0.00066984
5 0.00027614
};\addlegendentry{ }

\addplot[blue,mark=x,dashed,mark options=solid]
table[]{x y
1 0.0326
1.25 0.0306
1.5 0.0235
1.75 0.018409
2 0.014101
2.25 0.010889
2.5 0.0070119
2.75 0.0055698
3 0.003899
3.25 0.0023664
3.5 0.0012788
3.75 0.00076346
4 0.00037544
4.25 0.00021635
4.5 0.00011037
4.75 4.4507e-05
5 1.9044e-05
};\addlegendentry{ }

\addplot[blue,mark=x,dotted,mark options=solid]
table[]{x y
1 0.059348
1.25 0.064245
1.5 0.059736
1.75 0.056101
2 0.055479
2.25 0.058428
2.5 0.051935
2.75 0.059259
3 0.065876
3.25 0.061181
3.5 0.058292
3.75 0.061786
4 0.061143
4.25 0.06734
4.5 0.07391
4.75 0.066445
5 0.068966
};\addlegendentry{CRC + static 5G polar}

\addplot[brown,mark=+]
table[]{x y
1 0.1
1.25 0.1
1.5 0.1
1.75 0.1
2 0.1
2.25 0.1
2.5 0.1
2.75 0.1
3 0.1
3.25 0.1
3.5 0.1
3.75 0.1
4 0.1
4.25 0.1
4.5 0.1
};

\end{semilogyaxis}

\end{tikzpicture}
	\caption{BLER(solid), UER(dashed), MDR(dotted) vs. $E_b/N_0$ over the biAWGN channel for the $\left(64,42\right)$ dynamic RM code compared to a $(64,42+6)$ static 5G polar code with an outer CRC-$6$ \texttt{0x30}. SCL with $L=4$, threshold $\epsilon=0.1$}
	\label{fig:uBLER1}
\end{figure}

\begin{figure}[t]
	\centering
	\begin{tikzpicture}[scale=1]
\footnotesize
\begin{semilogyaxis}[
legend style={at={(0,0)},anchor= south west},
legend columns=3,
ymin=5e-6,
ymax=1,
width=3.5in,
height=3.5in,
grid=both,
xmin = 1,
xmax = 4.5,
xlabel = $E_b/N_0$ in dB,
ylabel = {BLER, UER, MDR},
]

\addplot[red,mark=o]
table[]{x y
1 0.83181
1.25 0.75206
1.5 0.65583
1.75 0.54948
2 0.43712
2.25 0.33114
2.5 0.23675
2.75 0.15819
3 0.10013
3.25 0.058968
3.5 0.032457
3.75 0.016789
4 0.0079574
4.25 0.0035826
4.5 0.0015206
};\addlegendentry{ }

\addplot[red,mark=o,dashed,mark options=solid]
table[]{x y
1 0.00024961
1.25 0.00034888
1.5 0.00037476
1.75 0.00044194
2 0.0004196
2.25 0.00043978
2.5 0.00038787
2.75 0.00026592
3 0.00019598
3.25 0.00013018
3.5 7.8429e-05
3.75 4.7605e-05
4 2.3646e-05
4.25 1.3181e-05
4.5 5.954e-06
};\addlegendentry{ }

\addplot[red,mark=o,dotted,mark options=solid]
table[]{x y
1 0.00030008
1.25 0.0004639
1.5 0.00057143
1.75 0.00080429
2 0.00095991
2.25 0.0013281
2.5 0.0016383
2.75 0.001681
3 0.0019572
3.25 0.0022076
3.5 0.0024164
3.75 0.0028355
4 0.0029716
4.25 0.0036792
4.5 0.0039155
};\addlegendentry{dynamic RM}

\addplot[blue,mark=x]
table[]{x y
1 0.76125
1.25 0.69868
1.5 0.6296
1.75 0.54965
2 0.46975
2.25 0.39054
2.5 0.30922
2.75 0.23665
3 0.17475
3.25 0.1222
3.5 0.081577
3.75 0.051548
4 0.030865
4.25 0.017414
4.5 0.0090606
};\addlegendentry{ }

\addplot[blue,mark=x,dashed,mark options=solid]
table[]{x y
1 0.002738
1.25 0.00248
1.5 0.0023776
1.75 0.0020125
2 0.001647
2.25 0.0014295
2.5 0.0012449
2.75 0.00089212
3 0.00067192
3.25 0.00044855
3.5 0.00031872
3.75 0.00024867
4 0.00014421
4.25 8.2004e-05
4.5 5.387e-05
};\addlegendentry{ }

\addplot[blue,mark=x,dotted,mark options=solid]
table[]{x y
1 0.0035967
1.25 0.0035496
1.5 0.0037764
1.75 0.0036614
2 0.0035062
2.25 0.0036603
2.5 0.0040259
2.75 0.0037698
3 0.003845
3.25 0.0036705
3.5 0.003907
3.75 0.004824
4 0.0046725
4.25 0.0047092
4.5 0.0059455
};\addlegendentry{CRC+ static 5G polar}

\addplot[brown,mark=+]
table[]{x y
1 0.005
1.25 0.005
1.5 0.005
1.75 0.005
2 0.005
2.25 0.005
2.5 0.005
2.75 0.005
3 0.005
3.25 0.005
3.5 0.005
3.75 0.005
4 0.005
4.25 0.005
4.5 0.005
};

\end{semilogyaxis}
   
\end{tikzpicture}
	\caption{BLER(solid), UER(dashed), MDR(dotted) vs. $E_b/N_0$ over the biAWGN channel for the $\left(64,42\right)$ dynamic RM code compared to a $(64,42+11)$ static 5G polar code with an outer CRC-$11$ \texttt{0x710}. SCL with $L=8$, threshold $\epsilon=0.005$}
	\label{fig:uBLER2}
\end{figure}
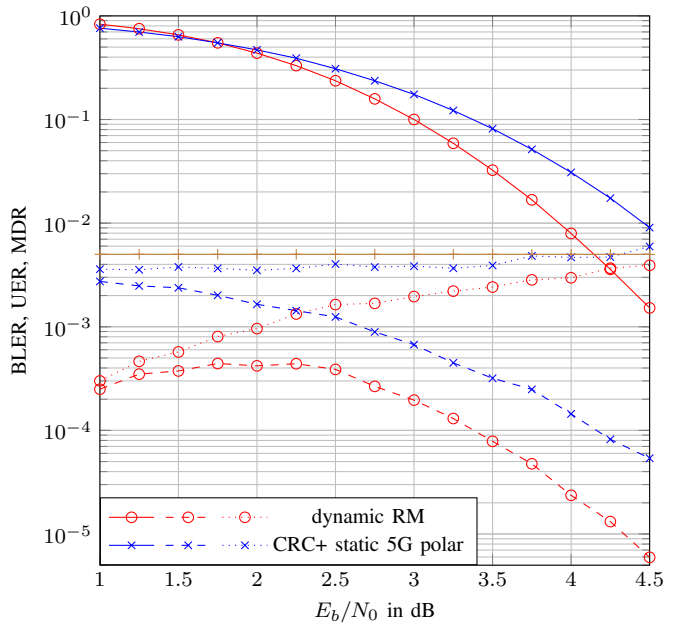

Fig.~\ref{fig:uBLER1} and Fig.~\ref{fig:uBLER2} show the \ac{BLER}, \ac{UER} and \ac{MDR} of the proposed decision rule~Eq.~\eqref{eq:decision_rule} based on the approximation~Eq.~\eqref{eq:APU} with \ac{SCL} decoding. For reference, we demonstrate the performance of CRC-concatenated polar codes using \ac{SCL} with the same list size. If none of the candidates in the list pass the CRC, an error flag is returned. Our method is compared with $6$ bits CRC using $L=4$ and $11$ bits CRC using $L=8$.\footnote{Note that the \ac{MDR} of CRC-concatenated polar codes using \ac{SCL} is influenced by both the CRC size and the list size. The generator polynomial is presented with Koopman's notation~\cite{koopman_crc}.} The threshold $\epsilon$ is chosen to achieve a similar \ac{MDR} to that of CRC-concatenated polar codes. The simulation results show that the proposed method outperforms CRC-concatenated polar codes in both \ac{BLER} and \ac{UER}. More importantly, the \ac{MDR} is restricted to not be higher than $\epsilon$ (the brown horizontal line in Fig.~\ref{fig:uBLER1} and Fig.~\ref{fig:uBLER2}).

\section{Some Applications}\label{sec:siso}
In addition to the primary application discussed in the previous section, the proposed algorithm shows promise of other potential applications. In this section, we explore several areas where the codebook probability can provide potential solutions.

\subsection{Polar-coded pilot-free channel estimation}
Pilot-free channel channel estimation based on frozen constraints is proposed in~\cite{yuan2021polar} and~\cite[Sec.6.2]{yuan_diss}. The \ac{CSI} $h$ is evaluated via 
\begin{align}\label{eq:yuan2021polar}
\hat{h} = \argmax_{h} Q_{U_\mathcal{F}|Y^N,H}\left(0^{N-K}\left|y^N,h\right.\right)
\end{align}
for polar-like codes with static frozen constraints. A related method for estimating \ac{CSI} utilizes the parity-check constraints of a \ac{LDPC} code~\cite{imad2010blind}. 

Base on the codebook probability given \ac{CSI},
\begin{align*}
Q_\mathcal{U}\left(y^N,h\right)\triangleq \sum_{u^N\in \mathcal{U}}Q_{U^N|Y^N,H}\left(u^N|y^N,h\right),
\end{align*}
the \ac{CSI} is determined by identifying a value that maximizes the codebook probability, i.e.,
\begin{align}
\hat{h} = \argmax_{h} Q_\mathcal{U}^*\left(y^N, h\right)
\end{align}
which is a generalization of Eq.~\eqref{eq:yuan2021polar}.

\subsection{List error rate prediction}
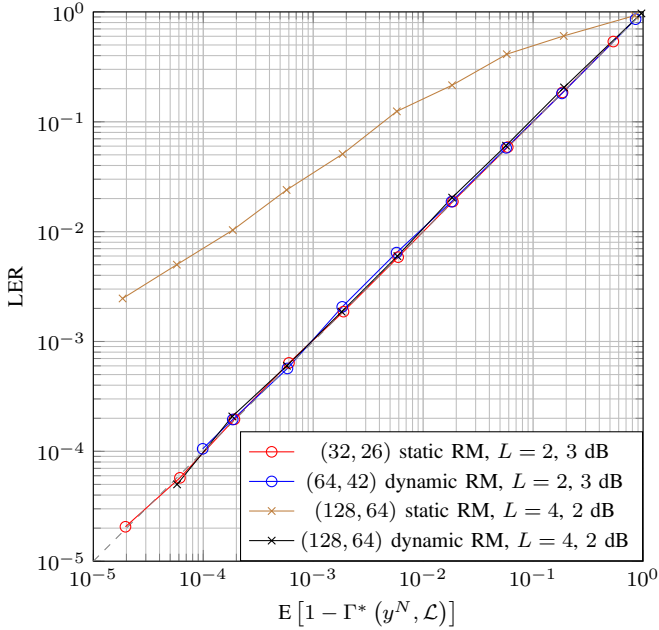
\begin{figure}[t]
	\centering
	\begin{tikzpicture}[scale=1]
\footnotesize
\begin{loglogaxis}[
legend style={at={(1,0)},anchor= south east},
ymin=1e-5,
ymax=1,
width=3.5in,
height=3.5in,
grid=both,
xmin = 1e-5,
xmax = 1,
xlabel = {$\text{E}\left[1-\Gamma^*\left(y^N, \mathcal{L}\right)\right]$},
ylabel = {LER},
]

\addplot[red,mark=o]
table[]{x y
0.53635 0.53709
0.18515 0.18492
0.058802 0.058778
0.018632 0.018788
0.0059225 0.0058592
0.0018859 0.0018698
0.00060166 0.00063854
0.00019206 0.00019593
6.1295e-05 5.7403e-05
1.9605e-05 2.059e-05
};\addlegendentry{$(32,26)$ static RM, $L=2$, $3$~dB}

\addplot[blue,mark=o]
table[]{x y
0.85523 0.85842
0.18286 0.18109
0.056852 0.0579
0.018099 0.018731
0.0057299 0.0064431
0.0018352 0.002066
0.0005831 0.0005685
0.00018532 0.00019491
9.8775e-05 0.00010545
};\addlegendentry{$(64,42)$ dynamic RM, $L=2$, $3$~dB}

\addplot[brown,mark=x]
table[]{x y
0.83899 0.92614
0.1899 0.60446
0.05761 0.41207
0.018317 0.21501
0.00576 0.12441
0.0018486 0.050847
0.00057252 0.023952
0.00018531 0.010322
5.7481e-05 0.005006
1.8464e-05 0.0024636
};\addlegendentry{$(128,64)$ static RM, $L=4$, $2$~dB}

\addplot[black,mark=x]
table[]{x y
0.96584 0.96859
0.19096 0.20541
0.056701 0.060627
0.018346 0.020407
0.0058301 0.0059846
0.0018149 0.001867
0.00057632 0.0006012
0.00018357 0.0002079
5.7767e-05 5.0021e-05
};\addlegendentry{$(128,64)$ dynamic RM, $L=4$, $2$~dB}

\addplot[gray,dashed]
table[]{x y
1e-5 1e-5
1 1
};

\end{loglogaxis}

\end{tikzpicture}
	\caption{Predicted \ac{LER} vs. simulated \ac{LER} of polar-like codes under SCL decoding with proposed scheme.}
	\label{fig:list_misdetection_rates}
\end{figure}

We evaluation the probability of \emph{the candidate list $\mathcal{L}_U$ contains the transmitted codeword} by generalizing Eq.~\eqref{eq:Gamma},
\begin{align}\label{eq:Gamma_list}
\Gamma^*\left(y^N, \mathcal{L}_U\right)
\triangleq \frac{\sum_{u^N\in\mathcal{L}_U}Q_{U^N|Y^N}\left(u^N|y^N\right)}{Q_\mathcal{U}^*\left(y^N\right)}.
\end{align}
Similar to section~\ref{sec:accuracy_block}, Fig.~\ref{fig:list_misdetection_rates} plots the \ac{LER} given the predicted \ac{LER}, where the \ac{LER} is defined as the probability of the transmitted codeword not being in the list. The results show that $\Gamma^*\left(y^N, \mathcal{L}_U\right)$ accurately predicts the \ac{LER} of the polar-like codes with dynamic frozen constraints.

\subsection{Bitwise soft-output decoding}
In various applications, the system requires post-decoding bitwise soft-output, e.g., iterative detection and decoding of \ac{MIMO} system, \ac{BICM-ID}, product codes and generalized \ac{LDPC} codes~\cite{Lentmaier10}. A \ac{SISO} decoder takes the sum of channel \acp{LLR} $\ell_i$, and a-priori \acp{LLR}, denoted as $\ell_{\text{A},i}$, as input,
\begin{align*}
\ell_i \triangleq\log\frac{p_{Y|C}(y_i|0)}{p_{Y|C}(y_i|1)},~
\ell_{\text{A},i} \triangleq \log\frac{P_{C_i}(0)}{P_{C_i}(1)},~i\in\left[N\right].
\end{align*}
It then outputs \ac{APP} \acp{LLR}, represented as $\ell_{\text{APP},i}$, and extrinsic \acp{LLR}, represented as $\ell_{\text{E},i}$.
\begin{align*}
\ell_{\text{APP},i} &\triangleq \log\frac{P_{C_i|Y^N}\left(0\left|y^N\right.\right) }{P_{C_i|Y^N}\left(1\left|y^N\right.\right) }\\
&= \log \frac{\sum_{c_i=0,c^N\in\mathcal{L}_C} P_{C^N|Y^N}\left(c^N\left|y^N\right.\right)}{\sum_{c_i=1,c^N\in\mathcal{L}_C} P_{C^N|Y^N}\left(c^N\left|y^N\right.\right)}\\
\ell_{\text{E},i} &\triangleq \ell_{\text{APP},i} - \ell_{\text{A},i} - \ell_i,~i\in\left[N\right].
\end{align*}

For an $(N, K)$ block code, the \ac{APP} \acp{LLR} can be determined using the \ac{BCJR} algorithm~\cite{bahl1974optimal} with $2^{N-K}$ states. Pyndiah's approximation~\cite{pyndiah_1998} extracts $\ell_{\text{APP},i}$ and $\ell_{\text{E},i}$ from a candidate list $\mathcal{L}_C$ through list decoding,
\begin{align}
&\ell_{\text{APP},i}^\prime=
      \log \frac{\max_{c_i=0,c^N\in\mathcal{L}_C} P_{C^N|Y^N}\left(c^N\left|y^N\right.\right)}{\max_{c_i=1,c^N\in\mathcal{L}_C} P_{C^N|Y^N}\left(c^N\left|y^N\right.\right)}\label{eq:pyndiah_app}\\
&\ell_{\text{E},i}^\prime=\left\{
    \begin{aligned}
      &+\beta, \text{ if } \left\{c_i=1,c^N\in\mathcal{L}_C\right\}=\varnothing\\
      &-\beta,\text{ if } \left\{c_i=0,c^N\in\mathcal{L}_C\right\}=\varnothing\\
      &\ell_{\text{APP},i}^\prime - \ell_{\text{A},i} - \ell_i, \text{ otherwise}.
    \end{aligned}\right.\label{eq:pyndiah_e}
\end{align}
Note that the saturation value $\beta>0$ is sensitive and requires optimization for practical applications.

By using the same approach as described in~\cite{yuan2023soft}, we approximate $\ell_{\text{APP},i}$ and $\ell_{\text{E},i}$ based on $Q_\mathcal{U}^*\left(y^N\right)$ via Eq.~\eqref{eq:approx_app}. In comparison to Pyndiah's approximation, Eq.~\eqref{eq:approx_app} introduces an additional term to dynamically adjust the weight between list observation and channel observation. Furthermore, Eq.~\eqref{eq:approx_app} eliminates the need for the saturation value present in Pyndiah's approximation. 

\begin{table*}[ht]
\begin{equation}\label{eq:approx_app}
\begin{aligned}
\ell_{\text{APP},i}\approx\ell^*_{\text{APP},i}\left(\mathcal{L}_U\right) &= \log \frac{\sum_{c_i=0,c^N\in\mathcal{L}_C} Q_{C^N|Y^N}\left(c^N\left|y^N\right.\right) +\left(Q_\mathcal{U}^*\left(y^N\right) - \sum_{c^N\in\mathcal{L}_C}Q_{C^N|Y^N}\left(c^N|y^N\right)\right)\cdot P_{C|Y}\left(0\left|y_i\right.\right)}{\sum_{c_i=1,c^N\in\mathcal{L}_C} Q_{C^N|Y^N}\left(c^N\left|y^N\right.\right)+ \left(Q_\mathcal{U}^*\left(y^N\right) - \sum_{c^N\in\mathcal{L}_C}Q_{C^N|Y^N}\left(c^N|y^N\right)\right)\cdot P_{C|Y}\left(1\left|y_i\right.\right)}\\
\ell_{\text{E},i}^* &= \ell_{\text{APP},i}^* - \ell_{\text{A},i} - \ell_i,~i\in\left[N\right]\\
&\text{where~} Q_{C^N|Y^N}\left(c^N\left|y^N\right.\right)=Q_{U^N|Y^N}\left(u^N\left|y^N\right.\right),~\text{for~}c^N=u^N\mathbb{F}^{\otimes \log_2 N}\\
&~~\text{and~} \mathcal{L}_C\triangleq \left\{c^N\in\left\{0,1\right\}^N: c^N=u^N\mathbb{F}^{\otimes \log_2 N}, \forall u^N\in\mathcal{L}_U \right\}.
\end{aligned}
\end{equation}
\end{table*}

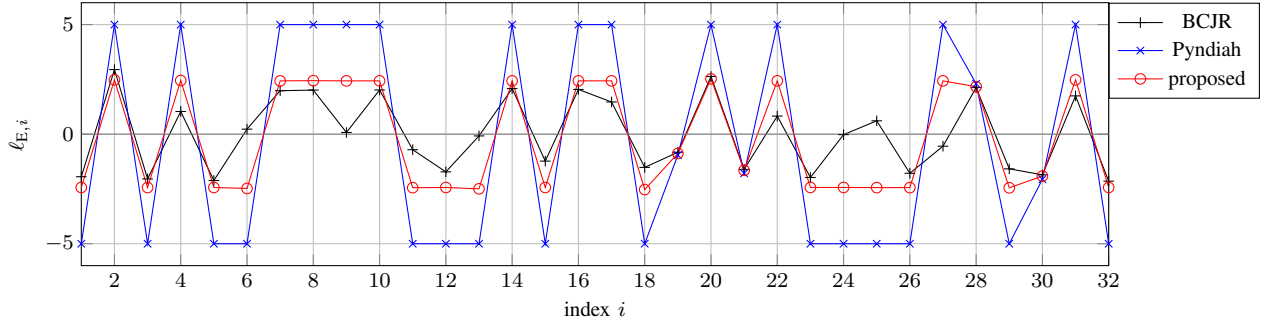
\begin{figure*}[t]
	\centering
	\begin{tikzpicture}[scale=1]
\footnotesize
\begin{axis}[
legend style={at={(1,1)},anchor= north west},
ymin=-6,
ymax=6,
width=6in,
height=2in,
grid=both,
xmin = 1,
xmax = 32,
xlabel = {index $i$},
ylabel = {$\ell_{\text{E},i}$},
]

\addplot[black,mark=+]
table[]{x y
1 -1.9424
2 2.947
3 -2.0417
4 1.0356
5 -2.1146
6 0.22936
7 1.9841
8 2.0123
9 0.077325
10 2.0175
11 -0.70875
12 -1.7199
13 -0.079753
14 2.0805
15 -1.2293
16 2.0418
17 1.4691
18 -1.5159
19 -0.82558
20 2.6323
21 -1.6074
22 0.82487
23 -1.9777
24 -0.022055
25 0.60991
26 -1.784
27 -0.54869
28 2.1247
29 -1.5831
30 -1.8549
31 1.7465
32 -2.1449
};\addlegendentry{BCJR}

\addplot[blue,mark=x]
table[]{x y
1 -5
2 5
3 -5
4 5
5 -5
6 -5
7 5
8 5
9 5
10 5
11 -5
12 -5
13 -5
14 5
15 -5
16 5
17 5
18 -5
19 -0.95523
20 5
21 -1.7855
22 5
23 -5
24 -5
25 -5
26 -5
27 5
28 2.3002
29 -5
30 -2.0589
31 5
32 -5
};\addlegendentry{Pyndiah}

\addplot[red,mark=o]
table[]{x y
1 -2.4313
2 2.4813
3 -2.4349
4 2.4498
5 -2.4321
6 -2.4777
7 2.4318
8 2.4423
9 2.4317
10 2.4334
11 -2.4422
12 -2.4332
13 -2.4949
14 2.4313
15 -2.4331
16 2.4321
17 2.4336
18 -2.5282
19 -0.88003
20 2.5251
21 -1.6514
22 2.4338
23 -2.4316
24 -2.4315
25 -2.4388
26 -2.4393
27 2.4326
28 2.1659
29 -2.4479
30 -1.9125
31 2.4784
32 -2.4316
};\addlegendentry{proposed}

\addplot[gray]
table[]{x y
1 0
32 0
};

\end{axis}

\end{tikzpicture}
	\caption{A random sample of the extrinsic information of a $(32,26)$ static RM code, $L=2$ at $E_b/N_0=2~\text{dB}$. In this example, the a-priori LLRs are set to $0$. The saturation value for Pyndiah's approximation is set to $\beta=5$.}
	\label{fig:L_E}
\end{figure*}

Fig.~\ref{fig:L_E} displays a random sample of the extrinsic LLRs $\ell_{\text{E},i}$ for a $(32,26)$ static RM code. Both Pyndiah's approximation and Eq.~\eqref{eq:approx_app} employ \ac{SCL} decoding with a list size of $2$. The saturation value $\beta$ for Pyndiah's approximation is set to $5$ in this example. We observe that Eq.~\eqref{eq:approx_app} provides extrinsic \acp{LLR} close to those obtained from the optimal BCJR algorithm and doesn't necessitate a pre-defined saturation value $\beta$.

\begin{figure}[t]
	\centering
	\begin{tikzpicture}[scale=1]
\footnotesize
\begin{semilogyaxis}[
legend style={at={(0,0)},anchor= south west},
ymin=3e-4,
ymax=2e-2,
width=3.5in,
height=3.5in,
grid=both,
xmin = 3,
xmax = 5,
xlabel = $E_b/N_0$ in dB,
ylabel = {BER},
]

\addplot[black,mark=+]
table[]{x y
3 0.011671
3.2 0.0089873
3.4 0.0066777
3.6 0.0051589
3.8 0.0039042
4 0.0026876
4.2 0.001892
4.4 0.0013328
4.6 0.00088785
4.8 0.00058381
5 0.00036464
};\addlegendentry{BCJR}

\addplot[red,mark=x]
table[]{x y
3 0.01431
3.2 0.011427
3.4 0.0086348
3.6 0.0067736
3.8 0.0053106
4 0.0039056
4.2 0.0027807
4.4 0.0019937
4.6 0.0013966
4.8 0.00093565
5 0.00061904
};\addlegendentry{proposed, $L=1$}

\addplot[red,mark=o]
table[]{x y
3 0.012119
3.2 0.0093266
3.4 0.0068846
3.6 0.0053066
3.8 0.0040193
4 0.0027957
4.2 0.0019532
4.4 0.0013863
4.6 0.00093112
4.8 0.00061102
5 0.00038114
};\addlegendentry{proposed, $L=2$}

\addplot[blue,mark=x]
table[]{x y
3 0.017518
3.2 0.014036
3.4 0.010621
3.6 0.0083987
3.8 0.0066853
4 0.0049889
4.2 0.0035518
4.4 0.0025447
4.6 0.0017957
4.8 0.0012321
5 0.00082527
};\addlegendentry{Pyndiah, $L=1$}

\addplot[blue,mark=o]
table[]{x y
3 0.012688
3.2 0.0098254
3.4 0.007251
3.6 0.0055806
3.8 0.0042042
4 0.0028995
4.2 0.0020405
4.4 0.0014481
4.6 0.0009583
4.8 0.00062605
5 0.00039158
};\addlegendentry{Pyndiah, $L=2$}

\end{semilogyaxis}

\end{tikzpicture}
	\caption{BER of $(32,26)$ static RM code under approximated \ac{APP} decoders and BCJR decoder.}
	\label{fig:BER}
\end{figure}
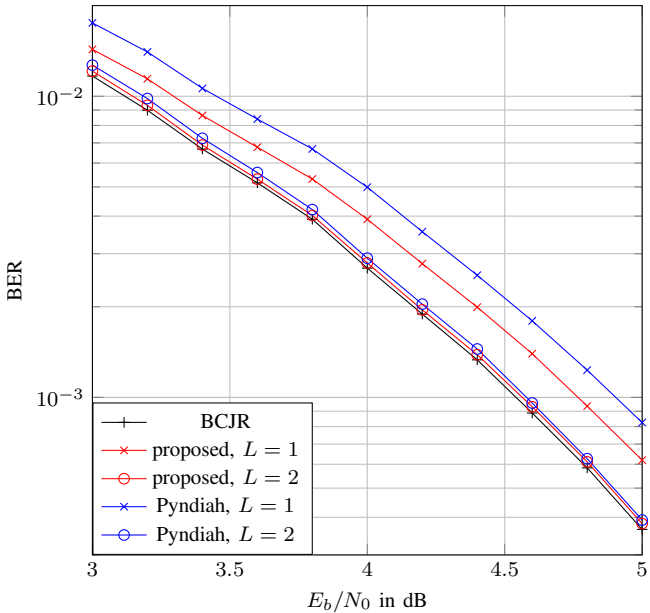
An \ac{APP} decoder, defined as
\begin{align*}
\hat{c}_i = \argmax_{a\in\{0,1\}} P_{C_i|Y^N}\left(a\left|y^N\right.\right)
\end{align*}
provides an optimal \ac{BER} by its definition. Fig.~\ref{fig:BER} shows that Eq.~\eqref{eq:approx_app} performs closer to the optimal BCJR decoder~\cite{bahl1974optimal} than Pyndiah's approximation Eq.~\eqref{eq:pyndiah_app}.

\subsection{Turbo product code decoding}

We consider $(N^2,K^2)$ product codes~\cite{elias_error-free_1954} based on static RM component codes. The block turbo decoding of product codes works as follows:
\begin{itemize}
    \item[0] The channel LLRs are stored in an $N\times N$ matrix $\mathbf{L}_\text{Ch}$. The a priori LLRs of the coded bits are initialized to all zero, i.e., $\mathbf{L}_\text{A}=\mathbf{0}$.
    \item[1] Each row of $\mathbf{L}_\text{Ch}+\mathbf{L}_\text{A}$ is processed by an SISO decoder, and the resulting APP LLRs and extrinsic LLRs are stored in the corresponding rows of $\mathbf{L}_\text{APP}$ and $\mathbf{L}_\text{E}$, respectively. A hard decision is made based on $\mathbf{L}_\text{APP}$. If all rows and columns of the decision correspond to valid codewords, the block turbo decoder returns the hard output, indicating successful decoding. Otherwise,  $\mathbf{L}_\text{A}$ is set to $\alpha\mathbf{L}_\text{E}$ and the decoder proceeds to the column update.
    \item[2] Each column of $\mathbf{L}_\text{Ch}+\mathbf{L}_\text{A}$ is decoded and the columns of $\mathbf{L}_\text{APP}$ and $\mathbf{L}_\text{E}$ are updated as step 1. A hard decision is performed using $\mathbf{L}_\text{APP}$. If the obtained binary matrix is valid, decoding success is declared. If the maximum iteration count is reached, a decoding failure is returned. Otherwise, we set $\mathbf{L}_\text{A} = \alpha\mathbf{L}_\text{E}$ and proceed to the next iteration (i.e., return to step 1).
\end{itemize}

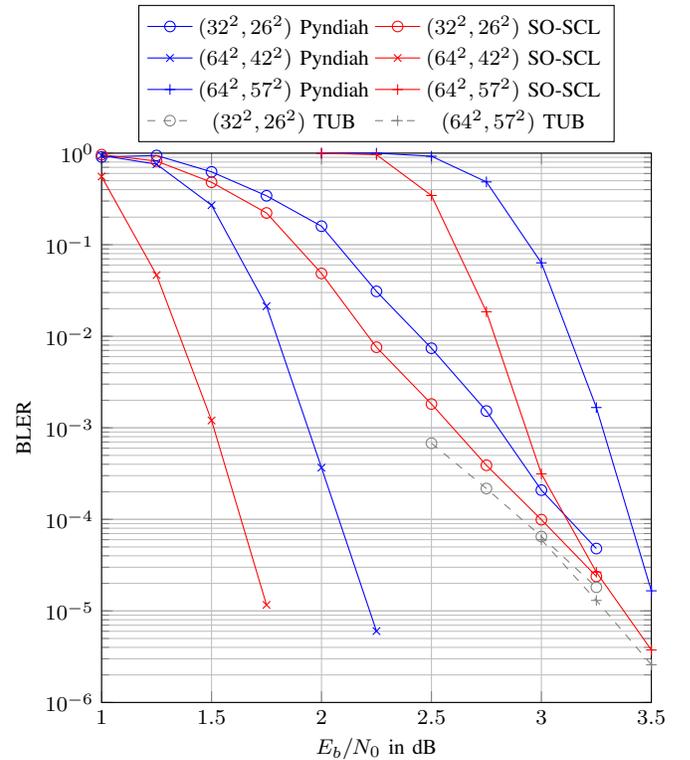
\begin{figure}[t]
	\centering
	\begin{tikzpicture}[scale=1]
\footnotesize
\begin{semilogyaxis}[
legend style={at={(0.5,1.02)},anchor= south},
legend columns=2,
ymin=1e-6,
ymax=1,
width=3.5in,
height=3.5in,
grid=both,
xmin = 1,
xmax = 3.5,
xlabel = $E_b/N_0$ in dB,
ylabel = {BLER},
]

\addplot[blue,mark=o]
table[]{x y
1 0.90909
1.25 0.9434
1.5 0.625
1.75 0.34247
2 0.15924
2.25 0.030921
2.5 0.0074008
2.75 0.0015251
3 0.00020853
3.25 4.8e-05
};\addlegendentry{$(32^2,26^2)$ Pyndiah}

\addplot[red,mark=o]
table[]{x y
1 0.96154
1.25 0.81967
1.5 0.48077
1.75 0.22222
2 0.048544
2.25 0.0075873
2.5 0.0018201
2.75 0.00038972
3 0.00009923
3.25 2.3825e-05
};\addlegendentry{$(32^2,26^2)$ SO-SCL}

\addplot[blue,mark=x]
table[]{x y
0 1
0.25 1
0.5 1
0.75 1
1 0.9434
1.25 0.75758
1.5 0.27027
1.75 0.021295
2 0.0003672
2.25 6.0208e-06
};\addlegendentry{$(64^2,42^2)$ Pyndiah}

\addplot[red,mark=x]
table[]{x y
0 1
0.25 1
0.5 1
0.75 0.90909
1 0.55556
1.25 0.046598
1.5 0.0012042
1.75 1.1625e-05
};\addlegendentry{$(64^2,42^2)$ SO-SCL}

\addplot[blue,mark=+]
table[]{x y
2 1
2.25 1
2.5 0.92593
2.75 0.48544
3 0.063371
3.25 0.0016652
3.5 1.6579e-05
};\addlegendentry{$(64^2,57^2)$ Pyndiah}

\addplot[red,mark=+]
table[]{x y
2 1
2.25 0.96154
2.5 0.34483
2.75 0.018525
3 0.00031403
3.25 2.6543e-05
3.5 3.7465e-06
};\addlegendentry{$(64^2,57^2)$ SO-SCL}

\addplot[gray,dashed,mark=o,mark options=solid]
table[]{x y
2.5 0.00067935
2.75 0.00021718
3 6.4992e-05
3.25 1.8136e-05
};\addlegendentry{$(32^2,26^2)$ TUB}

\addplot[gray,dashed,mark=+,mark options=solid]
table[]{x y
3 6.0028e-05
3.25 1.3021e-05
3.5 2.5841e-06
3.75 4.66711815555985e-07
};\addlegendentry{$(64^2,57^2)$ TUB}

\end{semilogyaxis}

\end{tikzpicture}
	\caption{BLER performance of the product codes with static RM component codes.}
	\label{fig:product_polar}
\end{figure}

Fig.~\ref{fig:product_polar} shows the comparison between the turbo decoder with Pyndiah's approximation~\cite{bioglio2019construction} Eq.~\eqref{eq:pyndiah_app}, Eq.~\eqref{eq:pyndiah_e}, and the proposed turbo decoder with \ac{SO}-SCL Eq.~\eqref{eq:approx_app}. Both turbo decoders have a maximum iteration number of $I_\text{max}=20$. All component codes are decoded by an \ac{SCL} decoder with list size of $L=4$ for both turbo decoders. For Pyndiah's decoder, $\alpha$ and $\beta$ parameters are taken from~\cite{pyndiah_1998}. The extrinsic \acp{LLR} are always scaled by $\alpha=0.5$ for the proposed turbo decoder. We demonstrate the performance of $(32^2,26^2)$, $(64^2,42^2)$ and $(64^2,57^2)$ product codes based on static RM component codes. The \acp{TUB} of $(32^2,26^2)$ and $(64^2,57^2)$ codes are provided. 

Simulation results show that the turbo decoder with proposed \ac{SO}-SCL significantly outperforms that with Pyndiah's approximation. For the high-rate $(32^2,26^2)$ and $(64^2,57^2)$ codes, the performance is close to their \acp{TUB} with \ac{SO}-SCL of list size $4$. 

Compared to turbo product code decoding with \ac{SOGRAND}~\cite{yuan2023soft}, SO-SCL has similar performance in \ac{BLER} for component codes with moderate $N-K$, resulting in product codes with similar dimensions to the LDPC codes used in 5G New Radio. SO-SCL can, however, decode component codes with large $N-K$ and thus can, additionally, decode long, extremely low rate product codes. However, SO-SCL requires a specific code structure due to its SC-nature, while SOGRAND is a universal decoder that is entirely agnostic to the code structure.

\section{Conclusions}\label{sec:conclusions}
In this work, we firstly proposed a metric, $\Gamma^*\left(y^N, \hat{u}^N\right)$, to evaluate the probability of the decision $\hat{u}^N$ being correct. The corresponding algorithm to compute the metric is compatible with all SC-based decoders. Our simulations confirm the accuracy of this metric in predicting the probability of correct decisions for polar-like codes with dynamic frozen bits. Utilizing this metric, we introduce a novel generalized decoding method. Simulation results highlight the superiority of dynamic \ac{RM} codes using this proposed approach over CRC-concatenated polar codes using \ac{SCL} decoding in both \ac{BLER} and \ac{UER}. Furthermore, we investigate potential applications of the codebook probability as coded pilot-free channel estimation, bitwise soft-output decoding and in turbo product code decoding.

\section{Acknowledgement}
This work was supported by the Defense Advanced Research Projects Agency under Grant HR00112120008.

\bibliographystyle{IEEEtran}
\bibliography{reference}

\end{document}